# Atomistic Line Graph Neural Network for Improved Materials Property Predictions

Kamal Choudhary[1,2,3,*], Brian DeCost[1,*]
1 Materials Measurement Laboratory, National Institute of Standards and Technology, Gaithersburg, MD, 20899, USA.
2 Theiss Research, La Jolla, California, 92037, USA.
3 DeepMaterials LLC, Silver Spring, MD, 20906, USA.
[*] These authors contributed equally to this work

**Abstract:**

Graph neural networks (GNN) have been shown to provide substantial performance improvements for atomistic material representation and modeling compared with descriptor-based machine learning models. While most existing GNN models for atomistic predictions are based on atomic distance information, they do not explicitly incorporate bond angles, which are critical for distinguishing many atomic structures. Furthermore, many material properties are known to be sensitive to slight changes in bond angles. We present an Atomistic Line Graph Neural Network (ALIGNN), a GNN architecture that performs message passing on both the interatomic bond graph and its line graph corresponding to bond angles. We demonstrate that angle information can be explicitly and efficiently included, leading to improved performance on multiple atomistic prediction tasks. We ALIGNN models for predicting 52 solid-state and molecular properties available in the JARVIS-DFT, Materials project, and QM9 databases. ALIGNN can outperform some previously reported GNN models on atomistic prediction tasks with better or comparable model training speed.

**Corresponding author**: Kamal Choudhary (kamal.choudhary@nist.gov)

## Introduction

Graphs are a powerful non-Euclidean data structure method for establishing relationships between features (nodes) and their relationships (edges) [1,2]. Graph neural networks (GNN)[3,4] have immense potential for modeling complex phenomena. Common applications of GNNs include community detection and link prediction in social networks[5,6], functional time series on brain structures[7], gene DNA on regulatory networks[8], information flow through telecommunications networks[9], and property prediction for molecular and solid materials[10]. From a quantum chemistry point of view, GNNs provide a unique opportunity to predict properties of solids, molecules, and proteins in a much faster way rather than by solving the computationally expensive Schrodinger equation[11-14]. There has been rapid progress in the development of GNN architectures for predicting material properties such as SchNet[10], Crystal Graph Convolutional Neural Networks (CGCNN)[15], MatErials Graph Network (MEGNet)[16], improved Crystal Graph Convolutional Neural Networks (iCGCNN)[17], OrbNet[18] and similar variants[19-31]. This family of models represents a molecule or crystalline material as a graph with one node for each constituent atom and edges corresponding to interatomic bonds. A common theme is the use of elemental properties as node features and interatomic distances and/or bond valences as edge features. Through multiple layers of graph convolution updating node features based on their local chemical environment, these models can implicitly represent many-body interactions. However, many important material properties (especially electronic properties such as band gaps) are highly sensitive to structural features such as bond angles and local geometric distortions. It is possible that these models are not able to efficiently learn the importance of such many-body interactions. Explicit inclusion of angle-based information has already been shown to improve models with hand-crafted features such as classical force-field inspired descriptors (CFID)[32]. Recently, there has been growing interest in the explicit incorporation of bond angles and other many-body features[17,20].

In this work, we use line graph neural networks inspired by those proposed in Ref[6] to develop an alternative way to include angular information to provide high accuracy models. Briefly, the line graph *L(g)* is a graph derived from another graph *g* that describes the connectivity of the edges in *g*. While the nodes of an atomistic graph correspond to atoms and its edges correspond to bonds, the nodes of an atomistic line graph correspond to interatomic bonds and its edges correspond to bond angles. Our model alternates between graph convolution on these two graphs, propagating bond angle information through interatomic bond representations to the atom-wise representations and vice versa. We use both the bond distances and angles in the line graph to incorporate finer details of atomic structure which leads to higher model performance. Our Atomistic Line Graph Neural Network (ALIGNN) models are implemented using the deep graph library (DGL) [33] which allows efficient construction and neural message passing for different types of graphs. ALIGNN is a part of the Joint Automated Repository for Various Integrated Simulations (JARVIS) infrastructure[34]. We train ALIGNN models for several crystalline material properties from JARVIS-density functional theory (DFT) [34-44] and Materials project[45] (MP) datasets as well as molecular properties from QM9[46] database.

## Results and discussion

### *A. Atomistic graph representation*

ALIGNN performs Edge-gated graph convolution[4] message passing updates on both the atomistic bond graph (atoms are nodes, bonds are edges) and its line graph (bonds are nodes, bond pairs with one common atom are edges). The Edge-gated graph convolution variant has the distinct advantage of updating both node and edge features. Because each edge in the bond graph directly corresponds to a node in the line graph, ALIGNN can aggregate features from bond pairs to efficiently update atom and bond representations by alternating between message passing updates on the bond graph and its line graph.

For crystals, we use a periodic 12-nearest-neighbor graph construction. We expand this nearest-neighbor graph to include edges to all atoms in the neighbor shell of the 12th-nearest neighbor. Each node in the atomistic graph is assigned 9 input node features based on its atomic species: electronegativity, group number, covalent radius, valence electrons, first ionization energy, electron affinity, block and atomic volume. This feature set is inspired by the CGCNN[15] model. The initial edge features are interatomic bond distances. We use a radial basis function (RBF) expansion with support between 0 and 8 Å for crystals and up to 5 Å for molecules. This undirected graph then can be represented as $G = (\upsilon, \epsilon)$ where υ are nodes and $\epsilon$ are edges i.e., a collection of $(\upsilon_i, \upsilon_j)$ linking vertices from $\upsilon_i$ to $\upsilon_j$. G has an associated node feature set $H=\{h_1,\ldots,h_N)$ , where $h_i$ is the feature vector associated with node $\upsilon_i$.

### *B. Atomistic line graph representation*

The atomistic line graph is derived from the atomistic graph. Each node in the line graph corresponds to an edge in the original atomistic graph; both entities represent interatomic bonds, and in our work, they share latent representations. Edges in the line graph correspond to triplets of atoms or pairs of interatomic bonds. The initial line graph edge features are an RBF expansion of the bond angle cosines: $\Theta = arccos\left(\frac{r_{ij} \cdot r_{jk}}{|r_{ij}||r_{jk}|}\right)$, where $r_{ij}$ and $r_{jk}$ are atomic displacement vectors between atoms *i, j,* and *k*. A schematic of an atomistic graph and corresponding atomistic line graph is shown in Fig. 1. To avoid ambiguity between the node and edge features of the atomistic graph and its line graph, we write atom, bond, and triplet representations as *h, e,* and *t*.

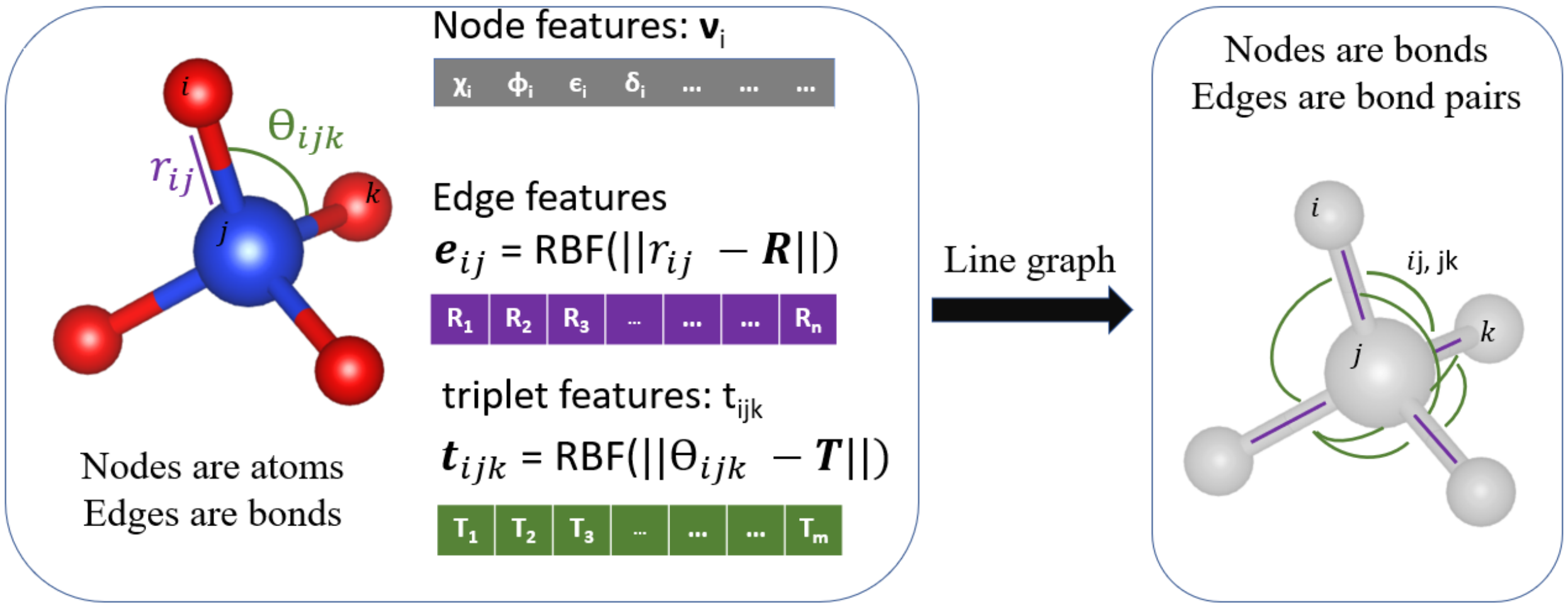

***Figure 1: Schematic showing undirected crystal graph representation and corresponding line graph construction for a $SiO_4$ polyhedron. For simplicity, only Si-O bonds are illustrated. The ALIGNN convolution layer alternates between message passing on the bond graph (left) and its line graph (or bond adjacency graph, right).***

### *C. Edge gated graph convolution*

ALIGNN uses Edge-gated graph convolution[4] convolution for updating both node and edge features. This convolution is similar to the CGCNN update, except that edge features are only incorporated into normalized edge gates. Furthermore, edge gated graph convolution uses the pre-aggregated edge messages to update the edge representations.

Edge gated graph convolution updates node representations $h^l$ from layer $l$ according to the formula:

$$h_i^{l+1} = f\left(h_i^l \{h_j^l\}_{j \in N_i}\right) \tag{1}$$

$$h_i^{l+1} = h_i^l + SiLU\left(Norm\left(W_{src}^l h_i^l + \sum_{j \in N_i} \hat{e}_{ij}^l \odot W_{dst}^l h_j^l\right)\right) \tag{2}$$

$$\hat{e}_{ij}^l = \frac{\sigma\left(e_{ij}^l\right)}{\sum_{k \in N_i} \sigma\left(e_{ik}^l\right) + \epsilon} \tag{3}$$

$$e_{ij}^{l} = e_{ij}^{l-1} + SiLU\left(Norm\left(A^{l}h_{i}^{l-1} + B^{l}h_{j}^{l-1} + C^{l}e_{ij}^{l-1}\right)\right) \quad (4)$$

The edge messages in this equation (4) are equivalent to the gating term in the CGCNN update[15], which coalesces the weight matrices *A*, *B*, and *C* into $W_{gate}$, and the augmented edge representation

$$z_{ij} = h_i \oplus h_j \oplus e_{ij} \quad (5)$$

$$e_{ij}^{l} = e_{ij}^{l-1} + SiLU\left(Norm\left(W_{gate}^{l} z_{ij}^{l-1}\right)\right) \quad (6)$$

***D. ALIGNN update***

One ALIGNN layer composes an edge-gated graph convolution on the bond graph (*g*) with an edge-gated graph convolution on the line graph (*L(g)*), as illustrated in Fig. 2. To avoid ambiguity between the node and edge features of the atomistic graph and its line graph, we write atom, bond, and triplet representations as *h*, *e*, and *t*. The line graph convolution produces bond messages *m* that are propagated to the atomistic graph, which further updates the bond features in combination with atom features *h*.

$$m^{l}, t^{l} = EdgeGatedGraphConv(L(g), e^{l-1}, t^{l-1}) \quad (7)$$

$$h^{l}, e^{l} = EdgeGatedGraphConv(g, h^{l-1}, m^{l}) \quad (8)$$

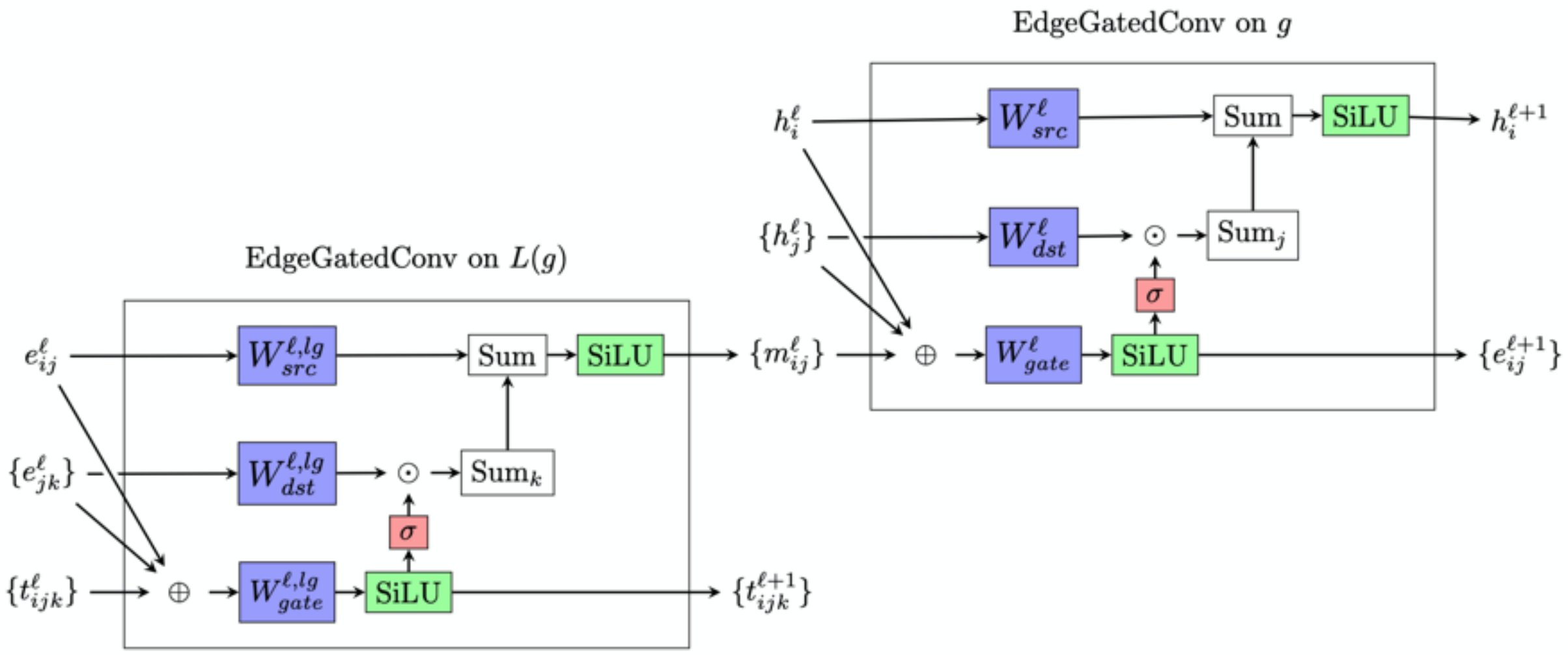

***Figure 2: Schematic of the ALIGNN layer structure. The ALIGNN layer first performs edge-gated graph convolution on the line graph to update pair and triplet features. The newly updated pair features are propagated to the edges of the direct graph and further updated with the atom features in a second edge-gated graph convolution applied to the direct graph.***

### *E. Overall model architecture and training*

We use $N$ layers of ALIGNN updates followed by $M$ layers of edge-gated graph convolution (GCN) updates on the bond graph. We use Sigmoid Linear Unit (SiLU, also known as Swish) activations instead of rectified linear unit (ReLU) or Softplus because it is twice differentiable like Softplus but can result in a better empirical performance like ReLU on many tasks. After $N + M$ graph convolution layers, our networks perform global average pooling over nodes and finally predict the target properties with a single fully connected regression or classification layers. Table 1 presents the default hyperparameters of the ALIGNN model used to train the models reported in Section F. These hyperparameters were selected through a combination of hypothesis-driven experiments and random hyperparameter search, as discussed in detail in the Methods section. Section G provides a detailed analysis of the sensitivity of model performance and computational cost.

***Table 1: ALIGNN model configuration used for both solid state and molecular machine learning models.***

| **Parameter** | **Value** |
| --- | --- |
| **ALIGNN layers** | 4 |
| **GCN layers** | 4 |
| **Edge input features** | 80 |
| **Triplet input features** | 40 |
| **Embedding features** | 64 |
| **Hidden features** | 256 |
| **Normalization** | Batch normalization |
| **Batch size** | 64 |
| **Learning rate** | 0.001 |

### *F. Model performance*

Model performance can vary substantially depending on the dataset and task. To evaluate the performance of ALIGNN, we currently use two different solid-state property datasets (Materials Project and JARVIS-DFT) as well as molecular property dataset QM9. Because the solid-state datasets are continuously updated, we use time-versioned snapshots of them, specifically selecting the MP version used by previous works to facilitate a direct comparison of model performance with the literature. It is likely that as these dataset sizes increase in the future the performance of the model can be further improved. We select the MP 2018.6.1 version which consists of 69239 materials with properties such as Perdew Burke-Ernzerhof functional (PBE)[47] bandgaps and formation energies. Similarly, we use 2021.8.18 version of JARVIS-DFT dataset, which consists of 55722 materials with several properties such as van der Waals correction with optimized Becke88 functional (OptB88vdW)[48] bandgaps, formation energies, dielectric constants, Tran-Blaha modified Becke Johnson potential (MBJ)[49] bandgaps and dielectric constants, bulk, shear modulus, magnetic moment, density functional perturbation theory (DFPT) based maximum piezoelectric coefficients, Boltztrap[50] based Seebeck coefficient, power factor, maximum absolute value of electric field gradient and two-dimensional materials exfoliation energies. All of these properties are critical for functional materials design. For the MP dataset we use a train-validation-test split of 60000-5000-4239 as used by SchNet[10] and MEGNet[16]. For the JARVIS-DFT dataset and its properties, we use 80 %:10 %: 10 % splits. For QM9 dataset we use a train-validation-test split of 110000-10000-10829 as used by SchNet[10], DimeNet++[20], and MEGNet[16].

Performance of ALIGNN models on MP is shown in Table 2, which shows the regression model performance in terms of mean absolute error metric (MAE). The best MAEs for formation energy ($E_f$) and band gap ($E_g$) with ALIGNN are 0.022 eV(atom)$^{-1}$ and 0.218 eV respectively. In terms of $E_f$, ALIGNN outperforms reported values of CGCNN, MEGNet and SchNet models by 43.6 %, 21.4 % and 37.1 % respectively. For $E_g$, ALIGNN outperforms CGCNN and MEGNet by 43.8 % and 33.9 % respectively. Good performance on well-known and well-characterized datasets ensures high prediction accuracy of ALIGNN models. Because each property has different units and in general a different variance, we also report the mean absolute deviation (MAD) for each property to facilitate an unbiased comparison of the model performance between different properties. The MAD values represent the performance of a random guessing model with average value prediction for each data point. We also report the CFID based predictions for comparison. Clearly, all the neural networks, especially ALIGNN, perform much better than the corresponding MAD of the dataset as well as CFID performance. Analyzing the MAD: MAE (ALIGNN) ratio, we observe that the ratio could be as high as 42.27 model. Generally, a model with high MAD:MAE ratio (such as 5 and above) is considered a good predictive model[51].

***Table 2: Test set performance on the Materials Project dataset. Predictions on test set are shown in parity plots in Supplementary Figure S1 and Supplementary Figure S2.***

| Prop | Unit | MAD | CFID | CGCNN | MEGNet | SchNet | ALIGNN | MAD: MAE |
|---|---|---|---|---|---|---|---|---|
| $E_f$ | eV(atom)$^{-1}$ | 0.93 | 0.104 | 0.039 | 0.028 | 0.035 | 0.022 | 42.27 |
| $E_g$ | eV | 1.35 | 0.434 | 0.388 | 0.33 | - | 0.218 | 6.19 |

Similarly, we train ALIGNN models on the JARVIS-DFT[34-44] dataset which consists of data for 55722 materials. In addition to properties such as formation energies, and bandgaps it also consists several unique quantities such as solar-cell efficiency (spectroscopic limited maximum efficiency, SLME), topological spin-orbit spillage, dielectric constant with ($\epsilon_x$ (DFPT)) and without ionic contributions ($\epsilon_x$ (OPT, MBJ)), exfoliation energies for two-dimensional (2D), electric field gradients (EFG), Voigt bulk (Kv) and shear modulus (Gv), energy above convex hull (ehull), maximum piezoelectric stress ($e_{ij}$) and strain ($d_{ij}$) tensors, n-type and p-type Seebeck coefficient and power factors (PF), crystallographic averages of electron ($m_e$) and hole ($m_h$) effective masses. As we converge plane wave-cutoff (ENCUT) and k-points used in Brillouin zone integration (Kpoint-length), we attempt to make machine learning predictions on these unique quantities as well. Such a large variety of properties allow a thorough testing of our ALIGNN models. More details for individual properties, its precision with respect to experimental measurements, applicability and limitations can be found in respective works. However, it is important to mention that many important issues such as tackling systematic underestimation of bandgaps by DFT methods, the inclusion of van der Waals bonding, and the inclusion of spin-orbit coupling interactions, all critically important for materials-design perspective have been key areas of improvements for the JARVIS-DFT dataset. For instance, meta-GGA (generalized gradient approximation) based Tran-Blaha modified Becke Johnson potential (TBmBJ) band gaps are more reliable and comparable to experimental data than Perdew Burke-Ernzerhof functional (PBE) or van der Waals correction with optimized Becke88 functional (OptB88vdW) bandgaps, but their calculations are computationally expensive and hence they are underrepresented in the dataset. In addition to the ALIGNN performance, we also include hand-crafted Classical force-field inspired descriptors (CFID) descriptor and CGCNN MAE performances for these properties using identical data-splits.

In Table. 3 we show the performance on regression models for different properties in the JARVIS-DFT database. We observe that ALIGNN models outperform CFID descriptors by up to 4 times, suggesting GNNs can be a very powerful method for multiple material property predictions. Also, ALIGNN outperforms CGCNN by more than 2 times (such as for OptB88vdW total energy).

Cross-dataset comparison of corresponding property entries in Table 2 and Table 3 shows that generally models generally obtain better performance on the MP dataset, which we attribute primarily to the larger size of MP. For example, the MAE for the formation energy target on MP dataset is 50% lower than for JARVIS-DFT. However, for some targets, the differences in the DFT method and settings, as well as potential differences in the material-space distribution, might significantly contribute to the difficulty of a prediction task. For example, the MAE on high throughput band gaps is lower (by 35.7 %) for the JARVIS-DFT dataset, which is interesting in light of MP's dataset size advantage over JARVIS-DFT. One potential source of this discrepancy is the differing computational methodologies used, such as different functionals (PBE vs OptB88vdW), use of the DFT+U method, and settings for various DFT hyperparameters like smearing and k-point settings, all of which can influence the values of computed bandgaps as discussed in Ref. [37]. Another potential contributing factor could be differing levels of dataset bias in the MP and JARVIS-DFT datasets stemming from differing distributions in material space. Clarifying this situation is beyond the scope of the present work, though it is of great importance for the atomistic modeling community to resolve.

Nevertheless, application of ALIGNN models on different datasets shows improvements for materials-property predictions. Both CFID, CGCNN and ALIGNN models' MAEs are lower than the corresponding MADs. The MAD:MAE ratios can vary for energy related quantities from a high value of 48.11 (total energy), and 26.06 (formation energy model) to low values such as for DFPT based piezoelectric strain coefficients (1.19) and dielectric constant with ionic contributions (1.63). The results indicate that there is still much room for improvement for the GNN models especially for electronic properties.

***Table 3: Regression model performances on JARVIS-DFT dataset for 29 properties using CFID, CGCNN and ALIGNN models on 55722 materials. Predictions on test set are shown in Supplementary Figure S3 to Supplementary Figure S31.***

| Property | Units | MAD | CFID | CGCNN | ALIGNN | MAD: MAE |
|---|---|---|---|---|---|---|
| Formation energy | $eV(atom)^{-1}$ | 0.86 | 0.14 | 0.063 | 0.033 | 26.06 |
| Bandgap (OPT) | eV | 0.99 | 0.30 | 0.20 | 0.14 | 7.07 |
| Total energy | $eV(atom)^{-1}$ | 1.78 | 0.24 | 0.078 | 0.037 | 48.11 |
| Ehull | eV | 1.14 | 0.22 | 0.17 | 0.076 | 15.00 |
| Bandgap (MBJ) | eV | 1.79 | 0.53 | 0.41 | 0.31 | 5.77 |
| Kv | GPa | 52.80 | 14.12 | 14.47 | 10.40 | 5.08 |
| Gv | GPa | 27.16 | 11.98 | 11.75 | 9.48 | 2.86 |
| Mag. mom | µB | 1.27 | 0.45 | 0.37 | 0.26 | 4.88 |
| SLME (%) | No unit | 10.93 | 6.22 | 5.66 | 4.52 | 2.42 |
| Spillage | No unit | 0.52 | 0.39 | 0.40 | 0.35 | 1.49 |
| Kpoint-length | Å | 17.88 | 9.68 | 10.60 | 9.51 | 1.88 |
| Plane-wave cutoff | eV | 260.4 | 139.4 | 151.0 | 133.8 | 1.95 |
| $\epsilon_x$ (OPT) | No unit | 57.40 | 24.83 | 27.17 | 20.40 | 2.81 |
| $\epsilon_y$ (OPT) | No unit | 57.54 | 25.03 | 26.62 | 19.99 | 2.88 |
| $\epsilon_z$ (OPT) | No unit | 56.03 | 24.77 | 25.69 | 19.57 | 2.86 |
| $\epsilon_x$ (MBJ) | No unit | 64.43 | 30.96 | 29.82 | 24.05 | 2.68 |
| $\epsilon_y$ (MBJ) | No unit | 64.55 | 29.89 | 30.11 | 23.65 | 2.73 |
| $\epsilon_z$ (MBJ) | No unit | 60.88 | 29.18 | 30.53 | 23.73 | 2.57 |
| $\epsilon$ (DFPT:elec +ionic) | No unit | 45.81 | 43.71 | 38.78 | 28.15 | 1.63 |
| Max. piezoelectric strain coeff ($d_{ij}$) | $CN^{-1}$ | 24.57 | 36.41 | 34.71 | 20.57 | 1.19 |
| Max. piezo. stress coeff ($e_{ij}$) | $Cm^{-2}$ | 0.26 | 0.23 | 0.19 | 0.147 | 1.77 |
| Exfoliation energy | $meV(atom)^{-1}$ | 62.63 | 63.31 | 50.0 | 51.42 | 1.22 |
| Max. EFG | $10^{21}$ $Vm^{-2}$ | 43.90 | 24.54 | 24.7 | 19.12 | 2.30 |
| avg. $m_e$ | electron mass unit | 0.22 | 0.14 | 0.12 | 0.085 | 2.59 |
| avg. $m_h$ | electron mass unit | 0.41 | 0.20 | 0.17 | 0.124 | 3.31 |
| n-Seebeck | $\mu VK^{-1}$ | 113.0 | 56.38 | 49.32 | 40.92 | 2.76 |
| n-PF | $\mu W(mK^2)^{-1}$ | 697.80 | 521.54 | 552.6 | 442.30 | 1.58 |
| p-Seebeck | $\mu VK^{-1}$ | 166.33 | 62.74 | 52.68 | 42.42 | 3.92 |
| p-PF | $\mu W(mK^2)^{-1}$ | 691.67 | 505.45 | 560.8 | 440.26 | 1.57 |

As we notice above, the regression tasks for some of the electronic properties do not show very high MAD: MAE. we train classification models for some of them. Classification tasks predict labels such as high value/low value (based on a selected threshold) as 1 and 0 instead of predicting actual data in regression tasks. Such models can be useful for fast screening purposes[38] for computationally expensive methods. We evaluate the performance of these classifiers using the receiver operating characteristic curve area under the curve (ROC AUC). A random guessing model has a ROC AUC of 0.5, while a perfect model would be a ROC AUC of 1.0. Interestingly, we notice most of our classification models (as shown in Table 4) have high ROC AUCs, ranging up to a maximum value of 0.94 (for convex hull stability) showing their usefulness for material classification-based applications. All results are based on the performance of 10 % test data which is never used during the training or model selection procedures.

***Table 4: Classification task ROC AUC performance on JARVIS-DFT dataset for ALIGNN models. The ROC curve plots for these models are provided in Supplementary Fig. S32 to Supplementary Fig.41.***

| **Model** | **Threshold** | **ALIGNN** |
|---|---|---|
| Metal/non-metal classifier (OPT) | 0.01 eV | 0.92 |
| Metal/non-metal classifier (MBJ) | 0.01 eV | 0.92 |
| Magnetic/non-Magnetic classifier | 0.05 µB | 0.91 |
| High/low SLME | 10 % | 0.83 |
| High/low spillage | 0.1 | 0.80 |
| Stable/unstable (ehull) | 0.1 eV | 0.94 |
| High/low-n-Seebeck | -100 $\mu VK^{-1}$ | 0.88 |
| High/low-p-Seebeck | 100 $\mu VK^{-1}$ | 0.92 |
| High/low-n-powerfactor | 1000 $\mu W(mK^2)^{-1}$ | 0.74 |
| High/low-p-powerfactor | 1000$\mu W(mK^2)^{-1}$ | 0.74 |

Next, we evaluate the ALIGNN model on QM9 molecular property dataset (130829 molecules) and compare it with other well-known models such as SchNet[10], MatErials Graph Network (MEGNet)[16], and DimeNet++[20] as shown in Table. 5. The results from models other than ALIGNN are reported as given in corresponding papers, not necessarily reproduced by us. QM9 provides DFT calculated molecular properties such as highest occupied molecular orbital (HOMO), lowest unoccupied molecular orbital (LUMO), energy gap, zero-point vibrational energy (ZPVE), dipole moment, isotropic polarizability, electronic spatial extent, internal energy at 0 K, internal energy at 298 K, enthalpy at 298 K, and Gibbs free energy at 298 K. ALIGNN outperforms competing methods for HOMO and dipole moment tasks while other accuracies are similar to the SchNet model. Most importantly, all ALIGNN results reported here use the same set of hyperparameters obtained by tuning to validation performance on the JARVIS-DFT bandgap target, suggesting that ALIGNN provides robust performance with respect to different datasets and material types.

***Table 5: Regression model performances on QM9 dataset for 11 properties using ALIGNN. These models were trained with same parameters as solid-state databases but for 1000 epochs. Predictions on test set are shown in Supplementary Figure S42 to Supplementary Figure S52.***

| **Target** | **Units** | **SchNet** | **MEGNet** | **DimeNet++** | **ALIGNN** |
|---|---|---|---|---|---|
| HOMO | eV | 0.041 | 0.043 | 0.0246 | 0.0214 |
| LUMO | eV | 0.034 | 0.044 | 0.0195 | 0.0195 |
| Gap | eV | 0.063 | 0.066 | 0.0326 | 0.0381 |
| ZPVE | eV | 0.0017 | 0.00143 | 0.00121 | 0.0031 |
| μ | Debye | 0.033 | 0.05 | 0.0297 | 0.0146 |
| α | $Bohr^3$ | 0.235 | 0.081 | 0.0435 | 0.0561 |
| $R^2$ | $Bohr^2$ | 0.073 | 0.302 | 0.331 | 0.5432 |
| U0 | eV | 0.014 | 0.012 | 0.00632 | 0.0153 |
| U | eV | 0.019 | 0.013 | 0.00628 | 0.0144 |
| H | eV | 0.014 | 0.012 | 0.00653 | 0.0147 |
| G | eV | 0.014 | 0.012 | 0.00756 | 0.0144 |

### *G. Model analysis*

We ablate individual components of the ALIGNN model to evaluate their contribution to the overall architecture. Keeping other parameters intact in the ALIGNN model (as specified in Table 1), we vary the number of ALIGNN and GCN layers as shown in Table 6 and Supplementary Table 1 for JARVIS-DFT OptB88vdW formation energies and bandgaps respectively. We find that without any graph convolution layers the MAE for the formation energy and bandgap are 1248.5 % and 453.6 % higher than the default model. Adding even a single ALIGNN or GCN layer can reduce the MAE by 102.9 % illustrating the importance of these layers. However, further increase in ALIGNN/GCN layers doesn't scale well and performance quickly saturates at a depth of 4. Excluding GCN layers and increasing ALIGNN layers and vice versa show the individual importance of these layers. Performance of GCN-only models saturates at 4 layers with 44 meV/atom MAE on the JARVIS-DFT formation energy task, while ALIGNN-only models saturate at 34 meV(atom)$^{-1}$—a relative reduction of 29.14 %. Each of these models, along with the other highlighted configurations in Table 6, performs four atom feature updates via graph convolution modules. At least two ALIGNN updates are needed to obtain peak performance. Additional atom feature updates provide little marginal increase in performance. This is consistent with the widely reported difficulty of GCN architectures scaling in depth beyond a few layers[52]. Figure 6 shows in detail the tradeoff between the performance benefit of including ALIGNN layers and their computational overhead relative to GCN layers. Per-epoch timing for each configuration is reported in Supplementary Table 2. All GCN-only configurations (annotated with the number of GCN layers) are on the low-computation portion of the pareto frontier, but the high-accuracy portion of the pareto frontier is dominated by ALIGNN/GCN combinations with at least two ALIGNN updates. The ALIGNN-2/GCN-2 configuration obtains peak performance (again, relative reduction of MAE by 29.14 %) with a computational overhead of roughly 2x relative to the GCN-4 configuration. Supplementary Table 1 and Supplementary Figure 53 present layer ablation study results yielding similar conclusions on the JARVIS-DFT OptB88vdW band gap target.

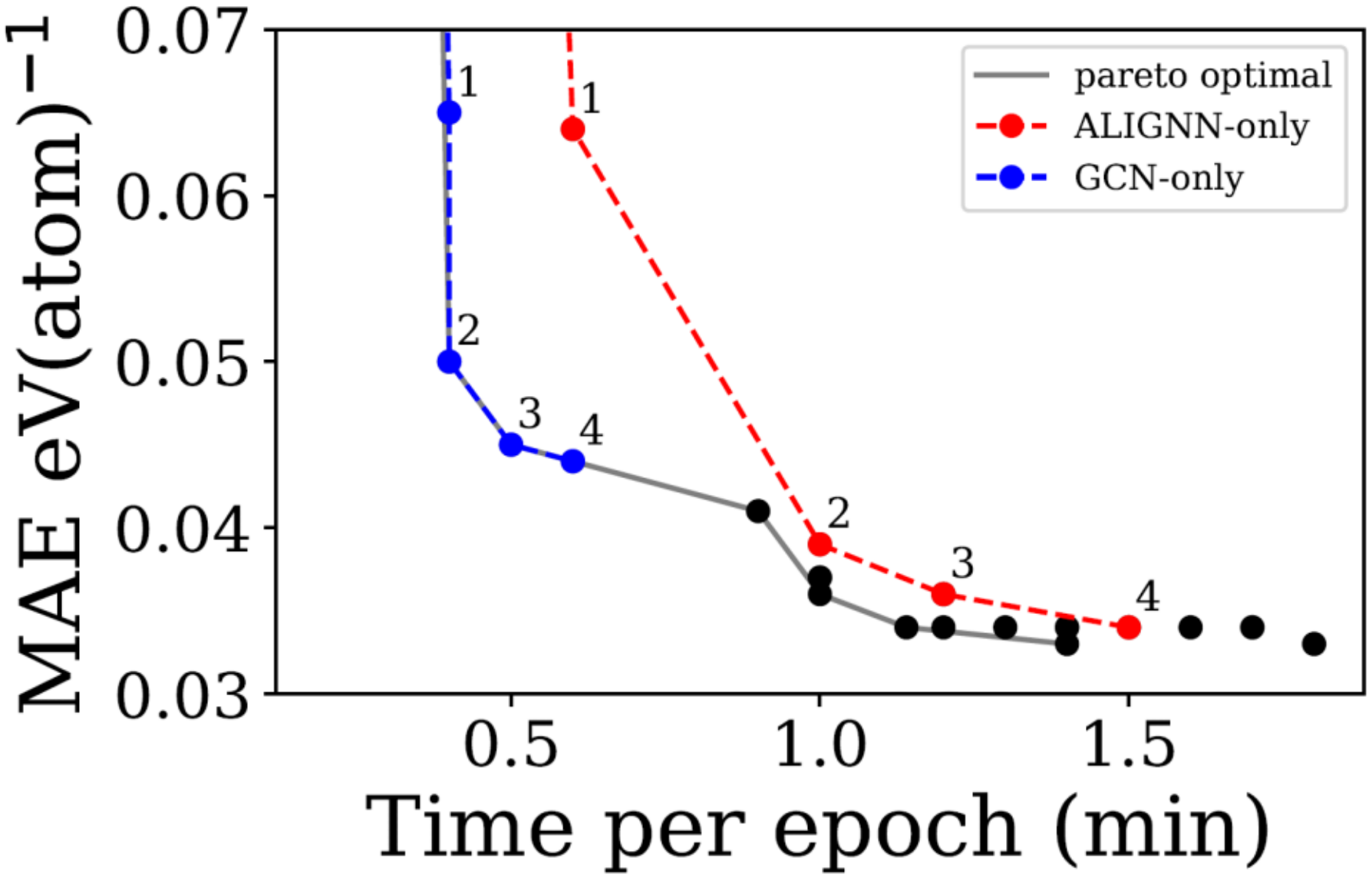

***Figure 3: ALIGNN layer study performance tradeoff, JARVIS-DFT formation energy target.***

This layer ablation study clearly demonstrates that inclusion of bond angle information and propagation of bond and pair features through the node updates improves the generalization ability of atomistic GCN models. This is satisfying from a materials science perspective, as interatomic bonding theory clearly motivates the notion that inclusion of bond angles should improve accuracy of the model.

Similarly, we vary the number of hidden features (*i.e.*, the width of the graph convolution layers), edge input features, and embedding input features to evaluate the MAE performance for JARVIS-DFT formation energy and bandgap model in comparison with the default model in Table 1. In Supplementary Table 3, we observe that the marginal performance from increasing the hidden features saturates at 256 for both properties. Supplementary Table 4 shows that the number of edge input features is optimal at 80 for formation energy model, while for the bandgap model performance saturates at 40. Similarly, embedding features are optimized at 64 for formation energy while 32 for bandgap model (Supplementary Table 5). Additionally, we tried three different node feature attributes such 1) CFID chemical features (total 438), only atomic number (total 1), and default CGCNN type attributes (total 92) and compared them for formation energy model in Supplementary Table 6. We observe that the default node attributes have the lowest MAE.

Next, we study time taken per epoch of several models for QM9 and JARVIS-DFT formation energy dataset in Supplementary Table 7. To help facilitate fair comparison, we train all models with the same computational resources using the reference implementations and configurations reported in the literature. We note that the timing code for the reference implementations of different methods may include differing amounts of overhead. For example, the ALIGNN timings reported in Supplementary Table 7 amortize the overhead of initial atomistic graph construction across 300 epochs, and each epoch includes the overhead of evaluating the model on the full training and validation sets for performance tracking. Additionally, the computational cost of deep learning models in general is not independent of certain hyperparameters; in particular, larger batch sizes can better leverage modern accelerator hardware by exposing more parallelism. We find ALIGNN requires less training time per epoch time compared to other models except DimeNet++ and MEGNet. However, it is important to note that DimeNet++ and other models usually take around 1000 epochs or more to reach desired accuracy, while ALIGNN can converge in about 300 epochs, resulting in lower overall training cost for similar or better accuracy.

While we report timing comparisons using our standard hyperparameter configuration used to train models reported in Section F, through subsequent model analysis we have identified several strategies that substantially reduce computational workload without incurring a large performance penalty. We observe in Supplementary Figure 54 that model performance converges after 300 epochs; shorter training budgets incur a modest performance reduction and slightly increased variance with respect to the training split. The performance tradeoff presented in Table 6 and Figure 3 indicates that switching from the default configuration of 4 layers each of ALIGNN and GCN updates to 2 layers each could offer a speedup of ~1.5x with negligible reduction in accuracy. Finally, we performed a drop-in replacement study comparing batch normalization and layer normalization in Supplementary Table 8, finding that switching to layer normalization provides an additional ~1.7x speedup with a slight degradation in validation loss and negligible degradation in validation MAE. Because the cost of retraining models for all targets reported is still high, and because some of these strategies equally apply to competing models, we defer a more comprehensive performance-cost study to future work.

Finally, we simultaneously investigate the effects of dataset size and different train-validation-test splits by performing a learning curve study in cross-validation for the JARVIS-DFT formation energy (Figure 4 and Supplementary Table 9) and bandgap (Supplementary Figure 55 and Supplementary Table 9) targets. We perform the cross-validation splitting procedure by merging the standard JARVIS-DFT train and validation sets and randomly sampling without replacement $N_{train}$ training samples and 5000 validation samples. The learning curve study shows no sign of diminishing marginal returns for additional data up to the full size of the JARVIS-DFT dataset. On the full training set size (44577) we obtain an average validation MAE of 0.0316 ± 0.0004 eV/at (uncertainty corresponds to the standard error of the mean over five cross-validation (CV) iterates). The standard deviation over CV iterates is 0.0009 eV/at, indicating that model performance is relatively insensitive to the dataset split.

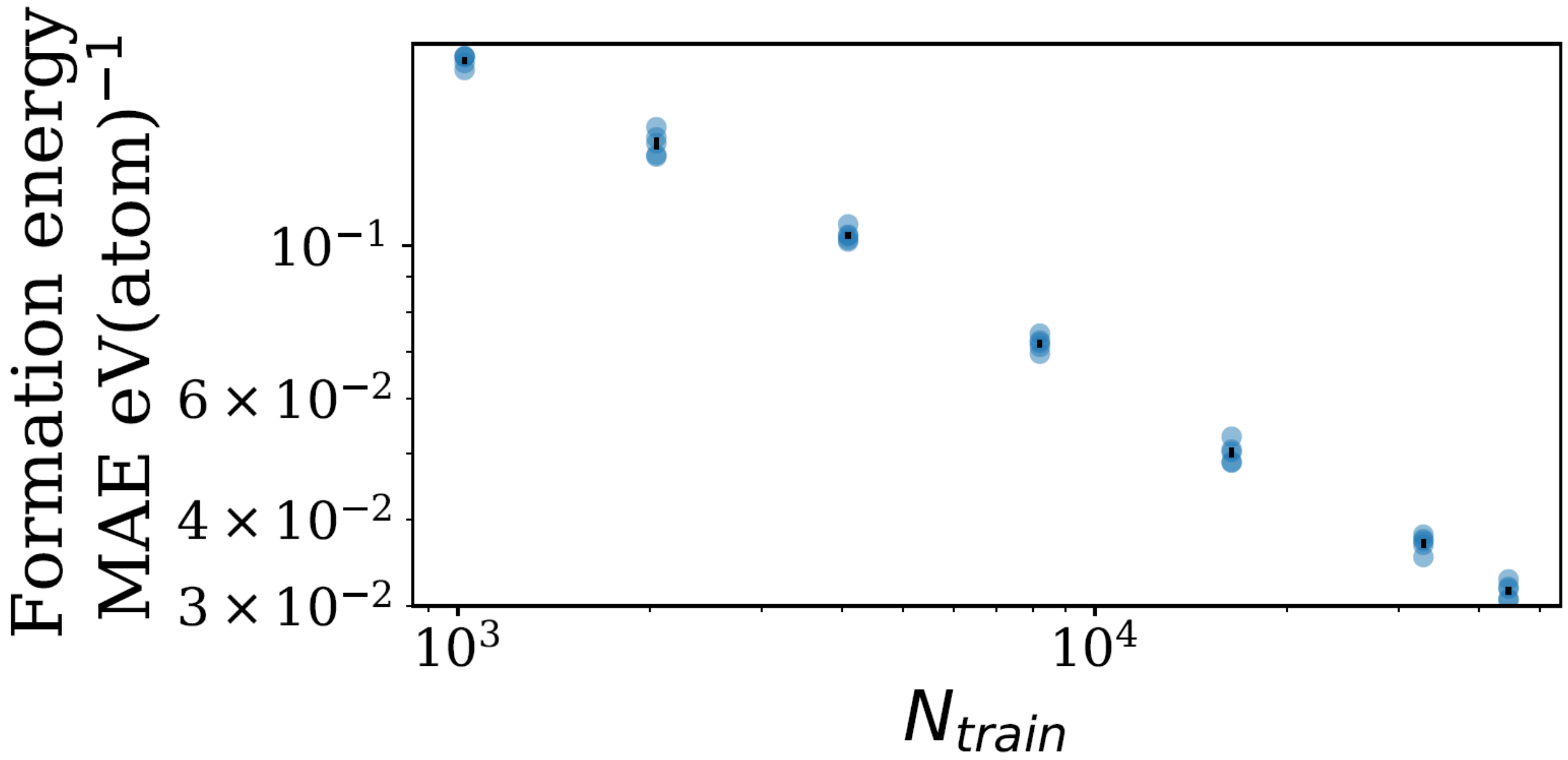

***Figure 4: learning curve for JARVIS-DFT formation energy regression target. Blue markers indicate validation set MAE scores for individual cross-validation iterates. Error bars indicate the mean cross-validation MAE +/- one standard error of the mean.***

In summary, we have developed an ALIGNN model which uses the line graph neural network that improves the performance of GNN predictions for solids and molecules. We have demonstrated that explicit inclusion of angle-based networks in GNNs can significantly improve model performance. A key contribution of this work is the inclusion and development of both the undirected atomistic graph and its line graph counterpart for solid-state and molecular materials. We develop regression and classification ALIGNN models for some of the well-known pre-existing databases and it can be easily applied for other datasets as well. Our model significantly improved accuracies over prior GNN models. We believe the ALIGNN model will rapidly improve the machine-learning prediction for several material properties and classes.

## Methods

### *A: JARVIS-DFT dataset*

The JARVIS-DFT[34-44] dataset is developed using Vienna Ab-initio simulation package (VASP)[53] software[54]. Most of the properties are calculated using the OptB88vdW functional[48]. For a subset of the data we use TBmBJ[49] for getting better band gaps. We use density functional perturbation theory (DFPT)[55] for predicting piezoelectric and dielectric constants with both electronic and ionic contributions. The linear response theory-based[56] frequency based dielectric function was calculated using both OptB88vdW and TBmBJ and the zero-energy values are trained for the machine learning model. Note that the linear response based dielectric constants lack ionic contributions. The TBmBJ frequency dependent dielectric functions are used to calculate the spectroscopic limited maximum efficiency (SLME)[38]. The magnetic moments are calculated using spin-polarized calculations considering only ferromagnetic initial configurations and neglecting any density functional theory (DFT)+U effects. The thermoelectric coefficients such as Seebeck coefficients and power factors are calculated using BoltzTrap[50] software using constant relaxation time approximation. Exfoliation energy for the van der Waals bonded two-dimensional materials are calculated as the energy per atom differences between the bulk and corresponding monolayer counterparts. The spin-orbit spillage[40] is calculated using the difference in wavefunctions of a material with and without inclusion of spin orbit coupling effects. All the JARVIS-DFT data and Classical force-field inspired descriptors (CFID)[32] are generated using the JARVIS-Tools package. The CFID baseline models are trained using the LightGBM package[54],[57] using the models developed in Ref.[32].

### *B. ALIGNN model implementation and training*

The ALIGNN model is implemented in PyTorch[58] and deep graph library (DGL)[33]; the training code heavily relies on PyTorch-ignite[59]. For regression targets we minimize the mean squared error (MSE) loss, and for classification targets we minimize the standard negative log likelihood loss. We train all models for 300 epochs using the AdamW[60] optimizer with normalized weight decay of $10^{-5}$ and a batch size of 64. The learning rate is scheduled according to the one-cycle policy[61] with a maximum learning rate of 0.001. We use the same model configuration for each regression and classification target. We use the initial atom representations from the CGCNN paper, 80 initial bond radial basis function (RBF) features, and 40 initial bond angle RBF features. The atom, bond, and bond angle feature embedding layers produce 64-dimensional inputs to the graph convolution layers. The main body of the network consists of 4 ALIGNN and 4 graph convolution (GCN) layers, each with hidden dimension 256. The final atom representations are reduced by atom-wise average pooling and mapped to regression or classification outputs by a single linear layer. These hyperparameters are selected to optimize validation MAE on the JARVIS-DFT band gap task through a combination of manual hypothesis-driven experiments and random hyperparameter search facilitated and scheduled through Ray Tune[62]; hyperparameter ranges are given in Supplementary Table 10. The random search results indicate that model performance is most highly sensitive to the learning rate, weight decay, and convolution layer width, and beyond a relatively low threshold is insensitive to the sizes of the initial feature embedding layers.

We used NIST's Nisaba cluster to train all ALIGNN models, and we reproduce results from the literature using the reference implementations for each competing method on the same hardware. Each model is trained on a single Tesla V100 SXM2 32 gigabyte Graphics processing unit (GPU), with 8 Intel Xeon E5-2698 v4 CPU cores for concurrently fetching and preprocessing batches of data during training[54]. For the MP dataset we use a train-validation-test split of 60000-5000-4239. For the JARVIS-DFT dataset, we use 80 %:10 %: 10 % splits. The 10 % test data is never used during training procedures. For QM9 dataset we use a train-validation-test split of 110000-10000-10829.

**Acknowledgements:** K.C. and B. D. thank the National Institute of Standards and Technology for funding, computational, and data management resources. Contributions from K.C. were supported by the financial assistance award 70NANB19H117 from the U.S. Department of Commerce, National Institute of Standards and Technology. This work was also supported by the Frontera supercomputer, National Science Foundation OAC-1818253, at the Texas Advanced Computing Center (TACC) at The University of Texas at Austin.

**Code availability:** The code and full model and training configurations used in this work are available on GitHub at https://github.com/usnistgov/alignn, along with general tooling at https://github.com/usnistgov/jarvis .

**Contributions:** Both K.C. and B.D. equally contributed to developing the model and writing the manuscript.

**Competing interests:**

The authors declare no competing interests.

**Data availability:** All data used in this work is available at Figshare link https://figshare.com/collections/ALIGNN_data/5429274. During the training these datasets are accessed using JARVIS-Tools's figshare module.

**Figure captions:**

Figure 1: Schematic showing undirected crystal graph representation and corresponding line graph construction for a $SiO_4$ polyhedron. For simplicity, only Si-O bonds are illustrated. The ALIGNN convolution layer alternates between message passing on the bond graph (left) and its line graph (or bond adjacency graph, right).

Figure 2: Schematic of the ALIGNN layer structure. The ALIGNN layer first performs edge-gated graph convolution on the line graph to update pair and triplet features. The newly updated pair features are propagated to the edges of the direct graph and further updated with the atom features in a second edge-gated graph convolution applied to the direct graph.

Figure 3: ALIGNN layer study performance tradeoff, JARVIS-DFT formation energy target.

Figure 4: learning curve for JARVIS-DFT formation energy regression target. Blue markers indicate validation set MAE scores for individual cross-validation iterates. Error bars indicate the mean cross-validation MAE +/- one standard error of the mean.